\documentstyle[11pt,emlines2]{article}
\def\bea{\begin{eqnarray}}
\def\eea{\end{eqnarray}}

\def\beq{\begin{equation}}
\def\eeq{\end{equation}}
\def\ba{\beq\new\begin{array}{c}}
\def\ea{\end{array}\eeq}
\def\be{\ba}
\def\ee{\ea}
\def\a{\alpha}
\def\b{\beta}
\def\g{\gamma}
\def\d{\delta}
\def\m{\mu}
\def\n{\nu}
\def\ep{\epsilon}
\makeatletter
\newdimen\normalarrayskip 
\newdimen\minarrayskip 
\normalarrayskip\baselineskip
\minarrayskip\jot
\newif\ifold \oldtrue \def\new{\oldfalse}

\def\arraymode{\ifold\relax\else\displaystyle\fi} 
\def\eqnumphantom{\phantom{(\theequation)}} 
\def\@arrayskip{\ifold\baselineskip\z@\lineskip\z@
\else
\baselineskip\minarrayskip\lineskip2\minarrayskip\fi}

\def\@arrayclassz{\ifcase \@lastchclass \@acolampacol \or
\@ampacol \or \or \or \@addamp \or
\@acolampacol \or \@firstampfalse \@acol \fi
\edef\@preamble{\@preamble
\ifcase \@chnum
\hfil$\relax\arraymode\@sharp$\hfil
\or $\relax\arraymode\@sharp$\hfil
\or \hfil$\relax\arraymode\@sharp$\fi}}

\def\@array[#1]#2{\setbox\@arstrutbox=\hbox{\vrule
height\arraystretch \ht\strutbox
depth\arraystretch \dp\strutbox
width\z@}\@mkpream{#2}\edef\@preamble{\halign
\noexpand\@halignto
\bgroup \tabskip\z@ \@arstrut \@preamble \tabskip\z@ \cr}
\let\@startpbox\@@startpbox \let\@endpbox\@@endpbox
\if #1t\vtop \else \if#1b\vbox \else \vcenter \fi\fi
\bgroup \let\par\relax
\let\@sharp##\let\protect\relax
\@arrayskip\@preamble}
\def\eqnarray{\stepcounter{equation}%
\let\@currentlabel=\theequation
\global\@eqnswtrue
\global\@eqcnt\z@
\tabskip\@centering
\let\\=\@eqncr
$$
\halign to \displaywidth\bgroup
\eqnumphantom\@eqnsel\hskip\@centering
$\displaystyle \tabskip\z@ {##}$%
\global\@eqcnt\@ne \hskip 2\arraycolsep

$\displaystyle\arraymode{##}$\hfil
\global\@eqcnt\tw@ \hskip 2\arraycolsep
$\displaystyle\tabskip\z@{##}$\hfil
\tabskip\@centering
&{##}\tabskip\z@\cr}
\begingroup\ifx\undefined\newsymbol \else\def\input#1 {\endgroup}\fi

\begin{document}

\setcounter{footnote}{1}
\def\thefootnote{\fnsymbol{footnote}}
\begin{center}
\hfill ITEP-TH-31/02\\ \hfill hep-th/0206191\\ \vspace{0.3in}
{\Large\bf      String theory derivation of RR couplings to D-branes}\\

\vspace{0.8cm}
\centerline{{\Large A.Ya.Dymarsky}\footnote{ ITEP and MSU, Moscow, Russia; e-mail: dymarsky@gate.itep.ru}}

\end{center}

\abstract{\footnotesize  }
\begin{center}
We derive the couplings of D-branes to the RR fields from
the first principles, i.e. from
the nonlinear $\sigma$-model.
We suggest a procedure to extract poles from string amplitudes
before the eventual formula for the couplings is obtained.
Using the fact that the parts  of amplitudes~ with  poles  are ~irrelevant ~for ~the~ effective
action this procedure ultimately simplify~
the~ derivation~ since it allows us to~ omit~ these~ poles already from the very beginning.
We also use the vertex for the RR fields in the  $(-{3\over 2},-{1 \over 2})$ ghost picture as
the useful tool for the calculation of such string amplitudes.
We carry out the calculations in all orders of brane massless excitations
and obtain the Myers-Chern-Simons action.
Our goal is to present the calculations in full technical details and
specify the approximation in which one obtains the action in question.
\end{center}

\bigskip
\setcounter{footnote}{0}
\renewcommand\thefootnote{\arabic{footnote}}

\section{Introduction}

\ \ \ The  couplings of the RR fields  to the D-branes have been   intensively studied recently from the various
points of view. In  several cases they were determined  using anomaly inflow
arguments \cite{Polch,Polch1,Green:1996dd,Li,Douglas}.
But this approach is inconsistent
 when transversal scalars are nonzero. In this case  the couplings  were derived using
properties of the string theory with respect to  T-duality transformations \cite{Myers} or
using some
mathematical tools  \cite{Emil}.  Unfortunately
when scalars are nonzero the interaction between the  RR fields and  excitations of branes  becomes more complicated,
but it still preserves   a nontrivial gauge  symmetry \cite{Alex}. Several of these results were
checked in the leading  orders by
calculating string-scattering amplitudes \cite{RR1,RR,Garo}.

In this paper we study  the couplings of RR fields to a stack of $N$~ D-branes in
the type    II superstring theory by calculating  string-scattering amplitudes.
We restrict ourself to the  tree-level amplitudes in the  low-energy limit.
We also assume that the only nontrivial fields are massless RR fields $C$,
 gauge fields $A_{\mu}$ and
scalar fields $\Phi^i$.

The main difficulty in this approach is the calculation
of integrals over locations of vertex operators.
This problem become more and more complicated when the number of vertexes
increases. However in order to determine the effective action we are looking
for only a part of all integrals is needed.
In fact, in order to determine the effective action for fields rather than
sources  one has to make the Legendre  transformation. 
In the case of tree-level  scattering amplitudes  this transformation
becomes simple --- one just has to remove all terms with poles in
momentum and leave finite in arbitrary momentum part.
If this extraction of singularities (poles in amplitudes) could be done {\bf before} the
calculation of integrals the  derivation of couplings  simplifies drastically.  Because most of  complicated, but irrelevant
expressions can be omitted.
In the next section we suggest the procedure of such  extraction
and discuss it's application in details.
Note that the integrals for the
finite parts of the amplitudes  are simple and can  be calculated for arbitrary number of vertex
operators. The result  is presented in appendix.

Before proceeding with   the general case with both nontrivial $A_{\mu}$ and $\Phi_i$ in section 4 we consider the simplified
case, when all matrix-valued scalars vanish. We use the RR vertex in $(-{1 \over2},-{1 \over2})$ picture
during this consideration. In order to compare our result with the
well-known one \cite{Polch} it is necessary to integrate by parts at the
end. In the general case nonzero scalar fields play the role of  transversal
coordinates and the RR field depends on them. That's why  integration by
parts and carrying  out other calculations is more complicated when scalars are nontrivial.
 We avoid these difficulties by using RR vertex in the $(-{3 \over2},-{1 \over2})$
picture, accordingly to the discussion in \cite{Polch}.
This approach allows  us to obtain low-energy effective action in all orders
of gauge and scalar fields.

The paper is organized as follows: in the next section we discuss
 the separation of the amplitudes into the finite part and the part with poles. In
 sections 3 and sections 4  we calculate the contribution from
the first part and see that it gives us the full  Myers \cite{Myers} action.
Section 5 ends the paper by conclusions.

\section{Separation of amplitudes}
Our aim now is to devide the string amplitudes, given by the following
$\sigma$-model correlator

\be \label{cft}
Tr\left\langle\int\limits_{\Sigma}V_{RR}~P\exp\left({\oint\limits_{\partial \Sigma}
V_{A,\Phi}}\right)\right\rangle
\ee
with vertexes $V_{RR}$ and $V_{A,\Phi}$   for RR and gauge/massless   matter
fields respectively, into two parts.  The first should contain finite terms (1PI graphs) and the second one --- all poles from 1PR graphs.

It is clear that poles in amplitudes arise when two vertex operators merge (come
close to each other). We could consider the  Wick pairing between those vertices, which
never merge. These pairings contribute  to finite (1PI) part and so to
the effective action.

In oder to avoid $SL_2$ divergence one should fix
the location of several vertex
 operators.
If we  fix $V_{RR}$
then there is a hole set of Wick pairings between $V_{RR}$ and all others,
which are finite in the  above sense.

We assume that besides those, finite at first glance terms, all others terms
contribute {\bf only}  to S-matrix poles and do not contain part finite in all
momentum. According to our assumption  we are going to consider only
Wick pairings between $V_{RR}$ and $V_{A,\Phi}$, but omit all pairings
between $V_{A,\Phi}$ themselves, because  all pairs of $V_{A,\Phi}$ meet together
during the integration over $\partial \Sigma$. We also conclude that the
separation should preserve $SL_2$ since we have already fixed it. In other
case the result would  depend on symmetry-fixing condition.

We will use the following  vertex operators: with ghost number
$-1$ in the next section
\be
\label{q}
V^{-1}_{A}(z)= A_{\mu}(P) c(z)e^{-\phi(z)}\psi^{\mu}(z)e^{2iP_{\mu}X^{\mu}(z)},~~~z\in \partial \Sigma
\ee
 and with ghost number $0$ in the section 4
\be
\label{w}
 V^{0}_{A,\Phi}(z)= c(z) ~e^{2iP_{\mu}X^{\mu}(z)}\left[^{\phantom{`^{\phantom{`}}}} A(P)_{\mu}
\partial X^{\mu}(z)+ {\alpha'}F_{\mu\nu}(P)\psi^{\mu}\psi^{\nu}(z) + \right. \cr
+\left.
\Phi(P) \partial
X^i(z)+i{\alpha'}[\Phi_i,\Phi_j](P)\psi^i\psi^j(z)+2{\alpha'}D_{\mu}\Phi_j(P)\psi^{\mu}\psi^j(z)^{\phantom{`^{\phantom{`}}}}
 \right] \ee
Indexes $\mu,\nu,...$
correspond to  dimensions along $D_p$-brane , $i,j,...$--- along
transversal  directions  and $A,B,...$--- along all. We  use doubling trick \cite{Kleb} in (\ref{q}),(\ref{w}) and
therefore all the fields on the boundary $\partial \Sigma$ depend on
$z$ but are independent on $\bar{z}$.
 Note that the conformal ghost anomaly on the disk is $-2$. Hence
to have non-zero correlation function the total ghost charge of vertex
operators in (\ref{cft}) should be 2.

It is important that vertices  (\ref{w}) contain  terms, which are nonlinear
in
external fields. These quadratic in $A_{\mu}$ and $\Phi^i$ expressions   are so-called contact terms \cite{contact}.
Their origin is the following. We are interested in the on-shell amplitudes. The correct way to obtain them is
to take from  (\ref{w}) only linear part and then calculate $\sigma$-model correlator  with arbitrary external
momentum. After this one should put the momentum on-shell. Unfortunately
carrying out this programm in all orders in  $A_{\mu}$ and $\Phi^i$  is
complicated and have never been done. But if one  adds contact terms to
vertexes before the calculation starts, one can calculate corelator with
on-shell momentum. These results should coincide. Thus contact terms are very
useful in simplification of our calculations.

The RR vertexes are  \cite{Polch,Kleb}
\be \label{RR} V^{(-{1\over 2},-{1\over 2})}_{RR}(w,\bar w)=\cr =\int
d^{10}P~
(P_{-}\widehat{\Im})^{\alpha\beta}(P):c(z)e^{-\phi(w)/2}e^{iP_{A}X^{A}(w)}S_{\alpha}(w):
:c(\bar{z})e^{-{\phi}(\bar w)/2}e^{iP_{A}\tilde{X}^{A}(\bar w)}{S}_{\beta}(\bar w):\cr
\widehat{\Im}^{\alpha\beta}(P)={i2\sqrt{2}\pi {\alpha'}\over 32 (p'+2)!}\int d^{10}X~
e^{-iP_{A}X^{A}}~
\Im_{A_0...A_{p'+1}}(X)~ \left[\Gamma^{A_0}...\Gamma^{A^{\phantom{1 \over 2}}_{p'+1}}\Gamma^{p+1}...\Gamma^9\Gamma^{11}\right]^{\alpha}_{\beta'} {\mathrm{C}}^{~\beta'\beta}
\ee
$$
 P_{\pm}=({\bf} 1 \pm \Gamma^{11}), ~~\Im_{A_0...A_{p'+1}}=(p'+2)\partial_{[A_0}C_{A_1...A_{p'+1}]},~~\tilde{X}^A=D^A_B X^B
$$
\be
 D^A_B= \left\{
  \begin{array}{l}
 \delta^A_B,~~~
 A\leq p
\\ -\delta^A_B,~~~ A>p
  \end{array} \right.
\ee
 and
\be \label{RR2}V^{(-{3 \over
2},-{1\over 2})}_{RR}(w,\bar w)=\cr =\int
d^{10}P~\widehat{C}^{\alpha\beta}(P):c(z)e^{-3\phi(w)/2}e^{iP_{A}X^{A}(w)/2}S_{\alpha}(w):
:c(\bar{z})e^{-{\phi}(\bar w)/2}e^{iP_A\tilde{X}^A(\bar
w)}{S}_{\beta}(\bar w):\cr
\widehat{C}^{\alpha\beta}(P)={i\sqrt{2}\pi \over 32 (p'+1)!}\int\limits d^{10} X~
e^{-iP_{A}X^{A}}~
 C_{A_0...A_{p'}}(X)~
\left [\Gamma^{A_0}...
\Gamma^{A^{\phantom{1 \over 2}}_{p'}}\Gamma^{p+1}...\Gamma^9\Gamma^{11}\right]^{\alpha}_{\beta'}{\mathrm{C}}^{\beta'\beta}.
\ee The coefficient ${i2\sqrt{2}\pi\alpha' \over 32 (p'+2)!}$  is taken for convenience here.
The origin of (\ref{RR2}) will be discussed in section 4.

Now we are going  to consider Wick pairing between $X$ fields.
 In the low-energy limit $P_{\mu}P^{\mu}~\rightarrow~0$
Wick pairings between $e^{2iP_{\nu}X^{\nu}(z)}$ are proportional to $|(z_1-z_2)|^{2\alpha' P_1 P_2}\rightarrow
1$ and  finite, even when vertexes merge.

The  pairing between $\partial_{\tau}X^{A}(z_1)e^{iP_{\nu}X^{\nu}(z_1)}$
 and $e^{iP_{\nu}X^{\nu}(z_2)}$ in the $P^2 \rightarrow~0$ limit
 goes to ${1\over z_1-z_2}$ and is singular when $z_1 \rightarrow z_2$.

Note that
there is a significant
 difference between  operators $A_{\mu}\partial X^{\mu}$ and $\Phi_i \partial X^i$ . In fact,
  operator $A_{\mu}\partial_t X^{\mu}$ could  be contracted with all other vertexes and $SL_2$ covariant expression is
 the sum over all these contractions (see example below). As we mentioned
above, among the latter there are divergent terms which we omit in our
calculations. So all the sum---all pairings should be omitted to respect $SL_2$.
 While $\Phi_i \partial_n X^i$ can be contracted
separately with $C_{A_0...A_{p'}}(X^{\mu},X^i)$, giving rise to
the following $SL_2$ covariant expression

\be \label{norm}
\left\langle
X^{i}(w,\bar{w})^{\phantom{1^{\phantom{1 }} }} \partial_n X^i(z,\bar{z}=z) \right\rangle =
 -{i \alpha'}g^{ij}{(w-\bar{w})\over (w-z)(\bar{w}-z)} \ee
That is why $\Phi_i\partial_n X^i$ should be taken into account
in the course  of our calculations.

Now recall  how to calculate the CFT correlator between
$:\psi\psi:$ and some others operators $O_k$ \cite{Zamol,Leht,Frie}:

\be
\label{psicor} \left\langle :\psi^{A}\psi^B (z):
:O^{a_1}_1(x_1):...:O^{a^{\phantom{1}}_n}_n(x_n):\right\rangle = \cr \sum_{m=1}^n
f_m(z,x_m)~\rho(M_{AB})^{a_m}_{b_m}\left\langle:O^{a_1}_1(x_1):...
:O_m^{a_m}(x_m):...:O^{a^{\phantom{1}}_n}_n(x_n):\right\rangle
\ee
Here $a_m$ denotes the index in
any representation of Lorentz group $O(9,1)$,
$\rho(M_{\mu\nu})^{a_m}_{b_m}$ is  the representation of the generator $M_{AB}$
 and $f_m(z,x_m)$ is a function corresponding to
operator $O_m$. For example, \be O=:\psi^C: ~~~~ f={1\over
z-x} ~~~~\rho(M_{\mu\nu})^C_D=\delta^B_D\cdot
g_{AC}-\delta^A_D\cdot g_{BC} - {\rm fundamental~~ representation} \ee
In the case of spin operator $S_{\alpha}$: \be O=S_{\alpha}~~~~~
f={1\over 2(z-x)}~~~~~
\rho(M_{\mu\nu})^{\alpha}_{\beta}={1\over
2}[\Gamma^{\mu},\Gamma^{\nu}]^{\alpha}_{\beta}-{\rm spinor~representation}
\ee

Formula (\ref{psicor}) should be  modified a little when
$O_m=:\psi^{C}\psi^{C'}:$. In fact, in this case

\be f={1 \over z-x}
\quad \rho(M_{\mu\nu})^{CC'}_{DD'}=(\delta^B_{D}\cdot
g_{AC}\delta^{C'}_{D'}-\delta^A_D\cdot g_{BC})+\delta^{C}_{D}(\delta^A_{D'}\cdot
g_{BC'}-\delta^B_{D'}\cdot g_{AC'})
\ee
($\rho$ is the generator
of the Lorentz group in the tensor square of the fundamental
representation) and we should also add \be \label{om} {1\over
(z-x_m)^2}(g^{AC'}g^{BC}-g^{AC}g^{BC'})
\left\langle
:O^{a_1}_1(x_1):...:\widehat{O}_m^{a_m}:...:O^{a^{\phantom{1^{\phantom{`}}}}_n}_n(x_n):\right\rangle
\ee to the RHS of (\ref{psicor}). Hat means that $O_m $ is omitted in  (\ref{om}).

If one wants to calculate  the correlator with more than one
$:\psi\psi:$ operator one  has to continue by induction: calculating step by
step, decreasing the number of  operators $:\psi\psi:$ inside the correlator
by one per one step. If the correlator is complicated it is
convenient to represent graphically all items in the
final sum. This technic is developed and discussed in \cite{Leht}.

To complete the evaluation of correlators like (\ref{psicor}) we
use the following results \cite{Leht,Frie}

\be \left\langle^{\phantom{1^{\phantom{1}}}} S_{\alpha}(w)~~ S_{\beta}(\bar{w}) ~~\right\rangle
= {\mathrm{C}}_{\alpha\beta}(w-\bar{w})^{-{5\over 4}}
\ee
and
\be
\left\langle^{\phantom{1^{\phantom{1}}}}S_{\beta}(\bar{w})~~ S_{\alpha}(w)~~\psi^{A}(z)  \right\rangle
= {1 \over \sqrt{2}}\Gamma^{\mu}_{\alpha\beta}(w-z)^{-{1\over
2}}(\bar{w}-z)^{-{1\over 2}}(w-\bar{w})^{-{3\over 4}}
\ee

Now we are ready to discuss the separation of fermion fields  correlators. We
start with  considering an example which can help us to realize
the specific features of $SL_2$ invariance of such  correlation
functions. An important point here is that both parts, finite one (1PI) and
the one with poles in momentum,  should be represented as the integrals of $SL_2$ covariant expressions.

 The example is (all vertexes are taken without $c$-ghost part here)
\be \label{ex}
\left\langle V^{-1}_{RR}(w,\bar{w},p)V^{-1}_{A}(z,k)V^{0}_{A}(x,k_1) \right\rangle = {-1
\over (w-z)(\bar{w}-z)(w-\bar{w})}\times \cr {1 \over \sqrt{2}}\left\{ -{i\sqrt{2}\alpha'}H^{\alpha\beta}(p)
A_{\mu}(k) A_{\nu}(k_1)\Gamma_{\alpha\beta}^{\mu} \left[{p^{\nu} \over
(x-w)}+{p^{\nu} \over (x-\bar{w})}+{2k^{\nu} \over (x-z)}\right] \right. + \cr
+{1\over 2} H^{\alpha\beta}(p)
A_{\mu}(k)F_{\sigma\rho}(k_1)\left[(\Sigma^{\sigma\rho}\Gamma^{\mu}{\mathrm{C}})_{\alpha\beta}{1
\over (x-w)}+ (\Gamma^{\mu}{\mathrm{C}}\Sigma^{\sigma\rho})_{\alpha\beta} {1
\over (x-\bar{w})} \right]+\cr \left.
+H^{\alpha\beta}(p)
A_{\mu}(k)F_{\sigma\rho}(k_1)\left[g^{\sigma\mu}\Gamma_{\alpha\beta}^{\rho}
- g^{\rho\mu}\Gamma_{\alpha\beta}^{\sigma}\right] {1 \over (x-z)}\right\},\cr
H^{\alpha\beta}=(P_{-})^{\alpha}_{\gamma}\hat{\Im}^{\gamma \beta} \quad \Gamma^{\mu}_{\alpha\beta}=(\Gamma^{\mu})_{\alpha}^{\gamma} {\mathrm{C}}_{\gamma\beta}, \quad
\Sigma^{A,B}={1\over2}[\Gamma^A,\Gamma^B], \quad p^2=k^2=k_1^2=0,~~p+k+k_1=0
\ee
We want now to separate eq. (\ref{ex}) into
two parts, preserving action of $SL_2$ group.
The first term is  covariant  due to on-shell
 condition $k_{\mu}A_{\mu}(k)=0$.
Second term  is also covariant due to
the following property of the gamma-matrixes

\be\label{13}
(\Sigma^{\sigma\rho})_{\alpha}^{\alpha'}(\Gamma^{\mu})_{\alpha'}^{\epsilon}{\mathrm{C}}_{\epsilon\beta}
=-(\Gamma^{\mu})_{\alpha}^{\epsilon}{\mathrm{C}}_{\epsilon\beta'}(\Sigma^{\sigma\rho})^{\beta'}_{\beta}
\ee
This property could be derived easily with the  convenient
choice of    matrix of charge conjugation
${\mathrm{C}}_{\alpha\beta}={\mathrm{C}}_{\beta\alpha} $ and
it's  main property ${\mathrm{C}}_{\a\g} (\Gamma^\mu)^{\g}_{\d} ({\mathrm{C}}^{-1})^{\d
\b}=(\Gamma^\mu)^{\b}_{\a}$\cite{Leht}.

In fact, we use that

\be
{\mathrm{C}}_{\a\g}(\Sigma^{\m\n})^\g_\b
={\mathrm{C}}_{\a\g}(\Gamma^{[\m})^\g_\d({\mathrm{C}}^{-1})^{\d\ep}{\mathrm{C}}_{\ep\varepsilon}(\Gamma^{\n]})^\varepsilon_\zeta
({\mathrm{C}}^{-1})^{\zeta\eta}{\mathrm{C}}_{\eta\b}=(\Gamma^{[\m})^\ep_\a
(\Gamma^{\n]})^\eta_\ep {\mathrm{C}}_{\eta\b}=-(\Sigma^{\m\n})_\a^\g {\mathrm{C}}_{\g\b}
\ee
and

\be
\label{e}
(\Gamma^{\m})^{\alpha}_{\epsilon}(\Sigma^{\sigma\rho})^{\epsilon}_{\beta}=(\Sigma^{\sigma\rho})_{\epsilon}^{\alpha}(\Gamma^{\m})^{\epsilon}_{\beta}
\ee
The  identity (\ref{e}) is true in the case, when $\m\neq\rho$ and  $\m\neq
\sigma$. This completes the derivation of (\ref{13}).

 The last term in (\ref{ex})
seems to be non-covariant. But if $\mu$ is equal either $\rho$ or $ \sigma$
then

\be
(\Sigma^{\sigma\rho}\Gamma^{\mu}{\mathrm{C}})_{\alpha\beta}+(\Gamma^{\mu}{\mathrm{C}}\Sigma^{\sigma\rho})_{\alpha\beta}\ee
is equal to
\be
-g^{\sigma\mu}\Gamma^{\rho}_{\alpha\beta}+g^{\rho\mu}\Gamma^{\sigma}_{\alpha\beta}
\ee
and together two last terms from (\ref{ex}) are covariant.

Thus, since the first and the last terms are singular when $x \to z$, by
assumption, only the second term  contributes to the effective action in question.
As well we also have obtained a nontrivial constraint that all three indexes $\mu,\rho,\sigma$ in the second term should be different.
This consideration could be easily generalized to arbitrary correlation
function in question.

In another words we see that separation of the correlators such as
(\ref{cft}) could be done in two steps. First, it is necessary to skip
$A_{\mu}\partial_{\tau} X^{\mu}$ from $V_{A,\Phi}^0$. Second, it is necessary
to consider Wick pairings (items in (\ref{psicor}), or graphs
in another words) only between $V_{RR}$ and any other
vertexes. As was described in the above example, we also have constraint that
all operators $\psi^A$ in vertexes, connected by Wick pairings, have
all distinct vector indexes.

\section{RR couplings to gauge fields only}

\ \ \ In this section we carry out all necessary calculations  when
matter fields $\Phi_i$ are absent and derive RR couplings
in this case.

We  use $V^{-1}_{RR}$ and one $V^{-1}_{A}$ in our calculus and preserve transformations $C \mapsto C+d\Lambda$
explicitly. Really, as mentioned in introduction $V^{-1}_{RR}$ depends
only on $\Im=dC$ and does not change under such transformations.
 On the other hand  the gauge
invariance under $A_{\mu}\mapsto A_{\mu}+ D_{\mu}\alpha$ is not
explicit in this approach until the eventual result is obtained.

Thus, we  change

\be P\exp \left(\oint \limits_{\partial \Sigma}
V^0_{A}\right)
\ee
from (\ref{cft}) to

\be \label{diff} 1 + \left\{ P
\sum_{n=0}^\infty ~~\oint \limits_{\partial \Sigma}V^{-1}_A
{1\over (n+1)!}\left(\oint \limits_{\partial
\Sigma}V^0_A\right)^n\right\} = 1 + \int_0^1 dt   ~P\left\{ \oint \limits_{\partial
\Sigma}V^{-1}_A
 \exp\left(\oint \limits_{\partial \Sigma} t V^{0}_A\right) \right\}
\ee
 in order not to change total ghost number of correlator.

First item --- 1 --- in this expression
corresponds to the fact that D-brane is the source for  the R-R field under
which it is charged:

\be
\label{cterm}
\int \limits d^{p+1 }X~ {1 \over (p+1)!} C_{\mu_0...\mu_p}\epsilon^{\mu_0...\mu_p}
\ee
This coupling is non-perturbative from the first-quantized string theory  point of
view \cite{Polch1}
and we will add it by hands in the end of our calculations. At this step we
remove 1 from (\ref{diff}).

Unfortunately the naive expression (\ref{diff}) does not preserve
SUSY. We see that all vertexes $V^0_{A}$ have been changed by multiplication by
$t$.   Recall that the contact terms are
  simply  the integrals of total
derivative from two-point correlation function of the linear  in $A_{\mu}$ vertexes~\cite{contact}.
Hence, if    $V_l=A_{\mu}(X)\partial_{\tau}X^{\mu}+
2\alpha'\partial_{[\mu}A_{\nu]}(X):\psi^{\mu}\psi^{\nu}:$ change into $tV_c $ then
$V_c=\alpha'[A_{\mu},A_{\nu}](x):\psi^{\mu}\psi^{\nu}:$ should become  proportional
to $t^2$. This means that correct contact term is $t^2 V_c$ and we change

\be
F_{\mu\nu}=\partial_\mu A_{\nu}-\partial_\nu
A_{\mu}+i[A_\mu,A_\nu]
\ee
from (\ref{w}) by

\be F^t_{\mu\nu}=t\partial_\mu A_{\nu}-t\partial_\nu
A_{\mu}+it^2[A_\mu,A_\nu]
\ee

Now the action in question is given by

\be \label{action} S=Tr \int_0^1 dt
\left\langle V^{-1}_{RR} P\left\{ V^{-1}_A
\exp\left(\oint_{\partial \Sigma} V^0_{A,t}\right) \right\}\right\rangle_{CFT}
\ee
where
\be
V^{0}_{A,t}=\alpha' F^t_{\mu\nu}(P)\psi^{\mu}\psi^{\nu}(z)  e^{2iP_{\mu}X^{\mu}(z)}
\ee
and we consider only that Wick
pairings in (\ref{action}), which are  shown in  graph~ü~1

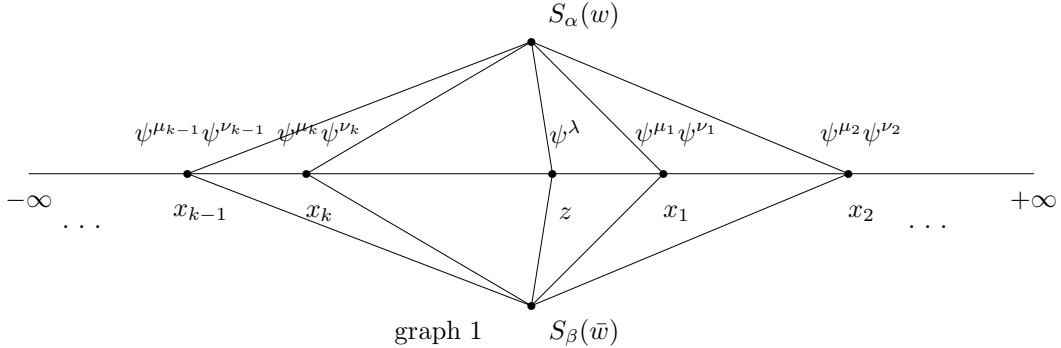
\begin{picture}(10,180.00)
\put(210,160){\makebox(0,0)[cc]{$S_{\alpha}(w)$}}
\put(210,40){\makebox(0,0)[cc]{$S_{\beta}(\bar{w})$}}

\put(155,40){\makebox(0,0)[cc]{graph~ü1}}

\put(0,90){\makebox(0,0)[cc]{$-\infty$}}
\put(380,90){\makebox(0,0)[cc]{$+\infty$}}
\put(340,80){\makebox(0,0)[cc]{$.~.~.$}}
\put(20,80){\makebox(0,0)[cc]{$.~.~.$}}

\put(65,85){\makebox(0,0)[cc]{$x_{k-1}$}}
\put(110,85){\makebox(0,0)[cc]{$x_k$}}
\put(203,85){\makebox(0,0)[cc]{$z$}}
\put(245,85){\makebox(0,0)[cc]{$x_1$}}
\put(315,85){\makebox(0,0)[cc]{$x_2$}}

\put(65,115){\makebox(0,0)[cc]{$ \psi^{\mu_{k-1}} \psi^{\nu_{k-1}} $} }
\put(110,115){\makebox(0,0)[cc]{$ \psi^{\mu_{k}} \psi^{\nu_{k}} $}}
\put(203,115){\makebox(0,0)[cc]{$ \psi^{\lambda} $}}
\put(245,115){\makebox(0,0)[cc]{$ \psi^{\mu_{1}} \psi^{\nu_{1}} $}}
\put(315,115){\makebox(0,0)[cc]{$ \psi^{\mu_{2}} \psi^{\nu_{2}} $}}

\emline{0}{100.00}{1}{380}{100.00}{2}

\put(60,100){\circle*{3}}
\put(105,100){\circle*{3}}
\put(198,100){\circle*{3}}
\put(240,100){\circle*{3}}
\put(310,100){\circle*{3}}

\put(190,50){\circle*{3}}
\put(190,150){\circle*{3}}

\emline{190}{50}{1}{60}{100}{1}
\emline{190}{150}{1}{60}{100}{1}

\emline{190}{50}{1}{105}{100}{1}
\emline{190}{150}{1}{105}{100}{1}

\emline{190}{50}{1}{198}{100}{1}
\emline{190}{150}{1}{198}{100}{1}

\emline{190}{50}{1}{240}{100}{1}
\emline{190}{150}{1}{240}{100}{1}

\emline{190}{50}{1}{310}{100}{1}
\emline{190}{150}{1}{310}{100}{1}
\end{picture}

\vspace{-1cm}

Before  our calculus starts note that P-ordering could be dropped.
This is so because traces over one matrix $A_{\lambda}$ and several matrixes
$F^t_{\mu_i\nu_i}$ in arbitrary order are equal to each other, if these
traces
are multiplied by  some tensor $T^{\lambda\mu_1\nu_1...}$ with property
\be
\label{bose}
T^{\lambda\mu_1\nu_1...\mu_k\nu_k...\mu_l\nu_l...}=T^{\lambda\mu_1\nu_1...\mu_l\nu_l...\mu_k\nu_k...}
\ee

In our case  tensor $T$ appears from the correlation function (\ref{psipsiSS}) (see below)
---the fermion functions correlator from (\ref{action}). The property
(\ref{bose})
is guaranteed by bose-symmetry.

\newpage
This leads us to the following expression for the action

\be
\label{action_2}
S= \int\limits d^{~p+1} X \sum_{k=0}^{\infty}{1 \over k!}
~\int_{-\infty}^{\infty}dx_1...\int_{-\infty}^{\infty} dx_k~
  \Im_{A_0...A_{p+1}}(X^{\mu},X^i=0) \times\cr
\times {i 2\sqrt{2}\pi \alpha' \over 32(p'+2)!}(w-z)(\bar{w}-z)(w-\bar{w})
{1\over(w-z)^{1\over 2}(\bar{w}-z)^{1\over 2}(w-\bar{w})^{1\over 4}}\times\cr
\times\left[P_{-}\Gamma^{A_0}...
\Gamma^{A^{\phantom{1^{\phantom{1}}}}_{p'+1}}\Gamma^{p+1}...\Gamma^9\Gamma^{11}\right]^{\alpha}_{\gamma}({\mathrm{C}}^{-1})^{\gamma\beta}
\left\langle S_{\alpha}(w)S_{\beta}(\bar{w})\psi^{\lambda^{
\phantom{`}}}(z):
\psi^{\mu_1}\psi^{\nu_1}(x_1):...:\psi^{\mu_k}\psi^{\nu_k}(x_k): \right \rangle_{gr.ü1}\cr
\times~(\alpha')^k \int_0^1 ~dt~ Tr ~\left[A_{\lambda}(X^{\mu})
F^t_{\mu_1 \nu_1}(X^{\mu})...F^{t^{\phantom{1}}}_{\mu_k\nu_k}(X^{\mu})\right],
\ee
The integral over D-brane world-volume appears from the integral over zero
modes of string coordinates in the functional integral for
(\ref{action}).
Here

\be \label{psipsiSS}
\left\langle S_{\alpha}(w)S_{\beta}(\bar{w})\psi^{\lambda^{
\phantom{`}}}(z):\psi^{\mu_1}\psi^{\nu_1}(x_1):
...:\psi^{\mu_k}\psi^{\nu_k}(x_k):\right\rangle_{gr.ü~1}=
{({\mathrm{C}}\Gamma^{\lambda})_{\alpha_0\beta_0}\over \sqrt{2}(w-z)^{1\over 2
}(\bar{w}-z)^{1\over 2}(w-\bar{w})^{3\over4}} \cr
\times{1\over
2}\left[{(\Sigma^{\mu_1\nu_1})^{\alpha_0}_{\alpha_1}\delta^{\beta_0}_{\beta_1}
\over (x_1-w)}+
{\delta^{\alpha_0}_{\alpha_1}(\Sigma^{\mu_1\nu_1})^{\beta_0}_{\beta_1}
\over (x_1-\bar{w})}\right]\times... \times{1\over
2}\left[{(\Sigma^{\mu_k\nu_k})^{\alpha_{k-1}}_{\alpha}\delta^{\beta_{k-1}}_{\beta}
\over (x_k-w)}+
{\delta^{\alpha_{k-1}}_{\alpha}(\Sigma^{\mu_k\nu_k})^{\beta_{k-1}}_{\beta}
\over (x_k-\bar{w})}\right]
\ee
In a full analogy with the example from section 2 the equal indexes in the set
$\mu_1,\nu_1,...,\mu_k,\nu_k,\lambda$ correspond to  singular expressions, when some of $x_i, z$ coincide.
Assuming that all $\mu_1,\nu_1,...,\mu_k,\nu_k,\lambda$ are distinct we
rewrite (\ref{psipsiSS}) as

\be {1\over
\sqrt{2}2^k}\left(\Gamma^{\lambda}\Sigma^{\mu_1\nu_1}...\Sigma^{\mu_k\nu_k}{\mathrm{C}}\right)_{\alpha\beta}
{(w-\bar{w})^k\over(w-z)^{1\over2}(\bar{w}-z)^{1\over2}(w-\bar{w})^{3\over4}
(x_1-w)(x_1-\bar{w})...(x_k-w)(x_k-\bar{w})}
\ee
Combining it with (\ref{action_2}) and performing the integration
over\footnote{This is done in appendix A.}
 $x_i$ one gets

\be
 \label{action3} S= \int\limits
d^{~p+1} X~\sum_{k=0}^{\infty}
\Im_{A_0...A_{p'+1}}(X^{\mu},X^i=0)\times\cr
\times{2\over 32 (p'+2)!} \left[P_{-}\Gamma^{A_0}...
\Gamma^{A^{
\phantom{`}}_{p+1}}\Gamma^{p+1}...\Gamma^9\Gamma^{11}\right]^{\alpha}_{\gamma}({\mathrm{C}}^{-1})^{\gamma\beta}
\left[\Gamma^{\lambda^{
\phantom{`}}}\Sigma^{\mu_1\nu_1}...\Sigma^{\mu_k\nu_k}{\mathrm{C}}\right]_{\alpha\beta}~\times\cr
\times {(i\pi\alpha')^{k+1}\over k!} ~Tr\int_0^1 dt ~A_{\lambda}(X^{\mu})
F^t_{\mu_1 \nu_1}(X^{\mu})...F^t_{\mu_k\nu_k}(X^{\mu})
\ee

The $\Gamma^{11}$ from $P_{-}$ in (\ref{action3}) has a simple
interpretation, due to identity
\be
{1\over(p'+2)!}\Gamma^{11}\Gamma^{A_0}...\Gamma^{A_{p'+1}}\Im_{A_0...A_{p'+1}}=
{1\over(9-p')!}\Gamma^{B_{p'+2}}...\Gamma^{B_9}\widetilde{\Im}_{B_{p'+2}...B_{9}}, \cr
*\Im=\widetilde{\Im}
\ee
This means that besides the action with the RR fields $C$ one gets the  action with dual RR fields $\widetilde{C}$ ($\widetilde{\Im}=d\widetilde{C}$).
In order to make calculus simple we remove $\Gamma^{11}$ from
(\ref{action3})
and  remember that the  action for dual RR
fields should be added at the end.

 Using the explicit
expression for the trace over gamma matrixes  one
finds that

\be\label{action4} S=\int_0^1 dt \int\limits
d^{p+1} X~\sum_{k=0}^{\infty}{2(i\pi \alpha')^{k+1}\over k!}(-1)^{p'}\delta_{p'+2+2k,p}\times \cr
\times\Im_{A_0...A_{p'+1}}(X^{\mu},X^i=0) ~Tr \left [A_{\lambda}(X^{\mu})
F^t_{\mu_1 \nu_1}(X^{\mu})...F^{t^{\phantom{`^{\phantom{`}}}}}_{\mu_k\nu_k}(X^{\mu})\right]\epsilon^{A_0...A_{p'+1}\lambda {\mu_1 \nu_1}...{\mu_k \nu_k}} \cr
= (-1)^{p'} (2 \pi i \alpha')\int_0^1 dt\int\limits  ~Tr \left[\Im\wedge A \wedge
e^{  2\pi i\alpha' F^t}\right]_{top}
\ee
 Here $\epsilon^{A_0...\nu_k}$ is a $(p+1)$--dimensional absolutely antisymmetric tensor.

Our result (\ref{action4}) has a remarkable structure: it is the
product of external derivative $\Im=dC$ and the  Chern-Simons term
\cite{Morozov}

\be
 CS^n=\int_0^1 dt ~Tr \left[A \wedge\underbrace{F^t\wedge...\wedge F^t}_{n~ times} \right]
\ee
 The Chern-Simons term is changed by full derivative under the gauge
 transformations $A_\mu\rightarrow A_\mu +D_{\mu} \alpha$ and this proves the gauge invariance of the action S.
 Its derivative is proportional to a gauge invariant Chern
 character  \cite{Morozov}

 \be
 d CS^n={1\over n+1} Tr \left[\underbrace{F\wedge...\wedge F}_{n+1~ times}\right]
 \ee
 After integration by parts the final answer is \cite{Polch,Myers} ($\lambda=2\pi i \alpha'$)

\be \label{action5} S= \int\limits ~Tr
\left(C \wedge e^{\lambda F} \right)_{top}
\ee

\section{RR couplings to both gauge and matter fields}
\subsection{RR vertex with $(-{3 \over 2},-{1\over 2})$ ghost number}
\textheight 25.0cm
In the end of previous section we showed that the use of $V^{-1}_{RR}$
leads to necessity to integrate by parts in the final expression (\ref{action4})-(\ref{action5}).
From the answer \cite{Myers}  we
know that when nonabelian fields $\Phi$ are present the RR field $C$
becomes the function of $\Phi^i$ rather than $X^i$ and
the integration by parts is much more complicated then. Thus in this
section instead of $V^{-1}_{RR}$~~(\ref{RR}) we will use $V^{-2}_{RR}$~(\ref{RR2}).
Vertex operator $V^{-2}_{RR}$ could be  constructed by the analogy with operator $V^{-1}_{RR}$
as the product of two spin vertices \cite{Frie,Knizh} (in the left and right sectors) with
defined ghost numbers\footnote {For similar consideration see \cite{Lerda} }.
Moreover $V^{-2}_{RR}$ and $V^{-1}_{RR}$ are connected by the picture-changing procedure.
 For further consideration it is convenient to represent $V_{RR}$ as
\be
V_{RR}^{(-{1 \over 2},-{1\over 2})}=(P_{-}\hat{\Im})^{\alpha\beta}V^{-{1\over
2}}_{\alpha}(w)V^{-{1\over
2}}_{\beta}(\bar{w}),\cr
V_{RR}^{(-{3 \over 2},-{1\over 2})}=\hat{C}^{\alpha\beta}V^{-{3\over
2}}_{\alpha}(w)V^{-{1\over
2}}_{\beta}(\bar{w}),\cr
V^{-{S\over
2}}_{\alpha}(\tilde{w})=c(\tilde{w})\widetilde{V}_{\alpha}^{-{S\over
2}}(\tilde{w})=c(\tilde{w})e^{-{S\over
2}\phi(\tilde{w})}S_{\alpha}(\tilde{w})e^{iP_A
\widehat{X}^{A}(\tilde{w})},\cr
~~\tilde{w}=w, \bar{w},~~\widehat{X}(w)=X(w),~~\widehat{X}(\bar{w})=\tilde{X}(\bar{w}),
\ee
The BRST operator \cite{Frie,belov} is given by

\be
Q_{BRST}=Q_0+Q_1+Q_2,\cr \ee
where $Q_0$ is built from the generators of the Virasoro algebra,
$Q_1$---from the
supersymmetry generator and $Q_2$ is needed for $Q^2_{BRST}=0$.
Since both $\widetilde{V}^{-{1\over
2}}_{\alpha}$ and $\widetilde{V}^{-{3\over
2}}_{\alpha}$ have conformal dimension  1, both commutators $[Q_0,V_{\alpha}^{-{S\over 2}}]=~0$.
It is also straightforward to see that commutators $[Q_2,V_{\alpha}^{-{S\over 2}}]$, for $S=1,3$ are equal to zero as well.
But \be
\left[Q_1,V_{\alpha}^{-{S\over 2}}(w)\right]={1\over \sqrt{2}}iP_A (\Gamma^A)_{\alpha}^{\beta}V_{\beta}^{1-{S\over
2}}\eta(w)\neq0\cr
\left[Q_1,V_{\alpha}^{-{S\over 2}}(\bar{w})\right]={1\over \sqrt{2}}iP_A D^A_B (\Gamma^B)_{\alpha}^{\beta}V_{\beta}^{1-{S\over
2}}\eta(\bar{w})\neq0
\ee
On-shell condition $d\Im=d*\Im=0$ leads to $P_A (\Gamma^A)^{\alpha}_{\gamma}\hat{\Im}^{\gamma\beta}=P_A
(\Gamma^A)^{\beta}_{\gamma}\hat{\Im}^{\alpha\gamma}=0$.
However for the $\hat{C}^{\alpha\beta}$ this is
not true, because $dC=\Im\neq0$ ($d*C=0$ is the gauge-fixing condition --- the analog of
$\partial_{\mu}A_{\mu}=0$). The consequence is that when
$V_{RR}^{(-{1\over2},-{1\over2})}$ is BRST closed the vertex
$V_{RR}^{(-{3\over2},-{1\over2})}$ is not closed.

It could be demonstrated by straightforward calculations, that
\be
\left[Q_{BRST},\xi V^{-{3\over 2}}_{\alpha}(w)\right]={1\over \sqrt{2}}iP_A (\Gamma^A)_{\alpha}^{\beta} V^{-{1\over 2}}_{\beta}(w)
\ee
when
\be
\left[Q_{BRST},\xi V^{-1}_{A}(z)\right]=V^{0}_A
\ee
where $\xi$ is the superconformal ghost.
Note that there is no $P_{-}$ in the definition for $V^{-2}$ and therefore  after $\xi$
manipulation \cite{Frie} in the left sector we obtain $V^{-1}$  with removed $P_{-}$. This
means that in the end of calculations with $V^{-2}$ one should simply add the
action for the dual RR fields. But this is not the end of the story. Since
$V^{-2}$ is not BRST closed correlation function with $V^{-2}$ and with $V^{-1}$
are not equal! However they are proportional to each other. More carefully \be
{1\over 2\sqrt{2}}\left<V_{RR}^{(-{1 \over 2},-{1\over 2})}(w,\bar{w}) V^{-1}_{A,\Phi}(z) \oint V^{0}_{A,\Phi}...\oint
V^{0^{\phantom{`}}}_{A,\Phi}\right>=\cr
{1\over 2\sqrt{2}} \hat{\Im}^{\alpha\beta}\int d\xi_0 \left<V_{\alpha}^{-{1 \over 2}}V_{\beta}^{-{1\over 2}}(w,\bar{w}) \xi(z)V^{-1}_{A,\Phi}(z) \oint V^{0}_{A,\Phi}...\oint
V^{0^{\phantom{`}}}_{A,\Phi}\right>=\cr
\hat{C}^{\alpha \beta} \int d\xi_0 \left<[Q_{BRTS},\xi V_{\alpha}^{-{3 \over 2}}(w)]V_{\beta}^{-{1\over 2}}(\bar{w}) \xi(z)V^{-1}_{A,\Phi}(z) \oint V^{0}_{A,\Phi}...\oint
V^{0^{\phantom{`}}}_{A,\Phi}\right>=\cr
=\hat{C}^{\alpha \beta}  \left< V_{\alpha}^{-{3 \over 2}}(w)V_{\beta}^{-{1\over 2}}(\bar{w}) V^{0}_{A,\Phi}(z) \oint
V^{0^{\phantom{`}}}_{A,\Phi}...\oint V^{0}_{A,\Phi}\right>+\cr
{1\over 2\sqrt{2}}\hat{C}^{\alpha \beta} iP_{A} (\Gamma^A)_{\beta}^{\gamma}\int d\xi_0 \left< :\xi V_{\alpha}^{-{3 \over 2}}(w)::\eta V_{\gamma}^{{1\over 2}}(\bar{w})::\xi V^{-1}_{A,\Phi}(z)::\oint
V^{0^{\phantom{`}}}_{A,\Phi}:...:\oint V^{0}_{A,\Phi}:\right>
\ee
Noting that
\be
\int d\xi_0 \left<:\xi(w)e^{-{3 \over 2}\phi(w)}::\eta(\bar{w})e^{{1 \over
2}\phi(\bar{w})}::\xi(z)e^{-\phi(z)}:\right>=\left<:e^{-{1 \over 2}\phi(w)}::e^{-{1
\over 2}\phi(\bar{w})}::e^{-\phi(z)}:\right>\cr
=(w-\bar{w})^{-{1 \over 4}}(w-z)^{-{1\over 2}}(\bar{w}-z)^{-{1\over 2}}
\ee
and
\be
\Gamma^{p+1}...\Gamma^{9}\Gamma^{11}\Gamma^{A}=(-1)^{p} D^A_B
\Gamma^B \Gamma^{p+1}...\Gamma^{9}\Gamma^{11}
\ee
we conclude that
\be
{1\over \sqrt{2}}\left<V_{RR}^{(-{1 \over 2},-{1\over 2})}(w,\bar{w}) V^{-1}_{A,\Phi}(z) \oint V^{0}_{A,\Phi}...\oint
V^{0}_{A,\Phi}\right>=\left<V_{RR}^{(-{3 \over 2},-{1\over 2})}(w,\bar{w}) V^{0}_{A,\Phi}(z) \oint V^{0}_{A,\Phi}...\oint
V^{0}_{A,\Phi}\right>
\ee
when $P_{-}$ removed from $V^{-1}$.
Therefore the action in question is given by
\textheight 23.0cm
\be
\label{_action} S=\sqrt{2}Tr \left\langle V^{-2}_{RR}~ P\exp \left(\oint_{\partial \Sigma} V^0_{A,\Phi}\right) \right\rangle_{CFT}
\ee
\subsection{The origin of the symmetric trace}

Let us explain now the appearance of the symmetrized trace $Str$
in the action we are looking for.
For a moment let us  consider a generic correlator

\be
\label{emil}
Tr\,P\,\int_{-\infty}^{\infty}dx_1...\int_{-\infty}^{\infty}dx_n
\left\langle V(x_1)^{\phantom{`^{\phantom{`}}}}...V(x_n)D(w)\right\rangle,\cr V=A a+B b+...+ Cc
\ee
Here $A,B,...,C$ are
the chan-paton matrixes and $a,b,...,c$ are the vertex operators
in $\sigma$-model. For any location of $x_1,...,x_n$ we have

\be
\label{STR} TrP \left\langle V(x_1)^{\phantom{`^{\phantom{`}}}}...V(x_n)D(w)\right\rangle = \cr Tr
\left\langle^{\phantom{`^{\phantom{`}}}}
\right(Aa+Bb+^{\phantom{`^{\phantom{`}}}}...+Cc\left)(x_{i_1})^{\phantom{`^{\phantom{`}}}}...\left(Aa+Bb+^{\phantom{`^{\phantom{`}}}}...+Cc\right)(x_{i_n})D(w)\right\rangle,\cr
x_{i_1}<...<x_{i_n},~~~~\forall k,k'\in \overline{1,n} ~~i_k \in
\overline{1,n},~~~x_k=x_{k'}\Rightarrow~~k=k' \ee Opening the brackets in
(\ref{STR}) and using the  symmetry of the correlation functions (at this
moment we assume it, but it is explicit in our case---see (\ref{_psipsiSS}) and
(\ref{CC}) below)

\be \left\langle^{\phantom{`^{\phantom{`}}}}...a(x_
{i})...b(x_{j})...\right\rangle = \left\langle^{\phantom{`^{\phantom{`}}}}...b(x_{i})...
a(x_{j})...\right\rangle
\ee
one could easily demonstrate that (\ref{STR}) is given by the sum
over all "words" of the length $n$ consisting of the "letters"
$a,b,...,c$ inside the correlator (in the arbitrary order)
multiplied by the sum of trace over the "words" also with length $n$
and made from "letters" $A,B,...,C$. But the $STr$ is exactly this
sum (divided by the numbers of such words) by definition. And one
could write (\ref{emil}) as

\be\label{STRR}
\int_{-\infty}^{\infty}dx_1...\int_{-\infty}^{\infty}dx_n
\sum_{l_a+...+l_c=n} {n!\over
{l_a!}...{l_c!}}STr\left[^{\phantom{`^{\phantom{`}}}}A(x_1)...A(x_{l_a})...C(x_{l_1})...C(x_{l_c})\right]\times
\cr \times \left\langle^{\phantom{`^{\phantom{`}}}} a(x_1)...a(x_{l_a})...c(x_{l_1})...c(x_{l_c})D(w)\right\rangle
\ee
Thus we see that $STr$ appears from the string-scattering
amplitudes in the natural way.

\subsection{Calculus}

Using the general result (\ref{STRR}) we present the action
(\ref{_action}) without the term (\ref{cterm}) in the form
\be
 \label{_action2}
 S=\sqrt{2}\sum_{k=1}^{\infty}\sum^{l+m+n+r=k}_{l,m,n,r}
\int\limits d^{p+1} X
~\int_{-\infty}^{\infty}dx_1...\int_{-\infty}^{\infty} dx_k\times
\cr \times {i \pi \sqrt{2}\over 32 (p'+1)!} 2^r ({\alpha'})^{m+n+r}
(x_1-w)(x_1-\bar{w})(w-\bar{w})\delta(x_1-z) {1
\over(w-\bar{w})^{3\over 4}}\times\cr
 \times ~{1 \over k!}{k!\over l!n!m!r!} \left\langle
C_{A_0...A_{p'}}(X^{\mu},X^i)\partial_n X^{i_1}(x_1)...\partial_n
X^{i_l}(x_l)\right\rangle
\left[\Gamma^{A_1}...\Gamma^{A^{\phantom{`}}_{p'}}\Gamma^{p+1}...\Gamma^9\Gamma^{11}\right]^{\alpha}_{\gamma}
({\mathrm{C}}^{-1})^{\gamma\beta}
\times\cr\times
\left\langle^{\phantom{`^{\phantom{`}}}} S_{\alpha}(w)S_{\beta}(\bar{w}):\psi^{\mu_1}\psi^{\nu_1}(x_1):
...:\psi^{\mu_n}\psi^{\nu_n}(x_l):
\cdot \right. \cr \left. \cdot
:\psi^{i_1}\psi^{j_1}(x_1):...:\psi^{i_m}\psi^{j_m}(x_m):
:\psi^{\lambda_1}\psi^{s_1}(x_1):...:\psi^{\lambda_r}\psi^{s_r}(x_r):^{\phantom{`^{\phantom{`}}}} \right\rangle_{gr.ü1}
\times\cr
\times STr ~ \left\{^{\phantom{`^{\phantom{`}}}} \Phi_{i_1}(X^{\mu})...\Phi_{i_l}(X^{\mu})
F_{\mu_1\nu_1}(X^{\mu})...F_{\mu_n\nu_n}(X^{\mu}) \cdot \right. \cr \cdot \left.
[\Phi_{i_1},\Phi_{j_1}](X^{\mu})...[\Phi_{i_m},\Phi_{j_m}](X^{\mu})
D_{\lambda_1}\Phi_{s_1}(X^{\mu})...D_{\lambda_r}\Phi_{s_r}(X^{\mu})^{\phantom{`^{\phantom{`}}}}\right \}
\ee

 Fermion correlator was already discussed and  here we  present only the
 result
\be \label{_psipsiSS}
\left\langle^{\phantom{`^{\phantom{`}}}}
S_{\alpha}(w)S_{\beta}(\bar{w}):\psi^{\mu_1}\psi^{\nu_1}(x_1):...:\psi^{\mu_n}\psi^{\nu_n}(x_n):
\cdot \right. \cr \left. \cdot
:\psi^{i_1}\psi^{j_1}(x_1):...:\psi^{i_m}\psi^{j_m}(x_m):
:\psi^{\lambda_1}\psi^{s_1}(x_1):...:\psi^{\lambda_r}\psi^{s_r}(x_r):^{\phantom{`^{\phantom{`}}}} \right\rangle_{gr.ü1}=\cr
 ={1\over
2^q}\left[\Sigma^{\mu_1\nu_1}...\Sigma^{\mu_n\nu_n} \Sigma^{i_1 j
_1}...\Sigma^{i_m j_m} \Sigma^{\lambda_1 s_1}...\Sigma^{\lambda_r
s_r}C\right]_{\alpha\beta}\times \cr \times {(w-\bar{w})^q\over
(w-\bar{w})^{5 \over 4}
(x_1-w)(x_1-\bar{w})...(x_q-w)(x_q-\bar{w})},~~ q=m+n+r \ee

Correlator between $X^i$ is also easy to calculate using Wick's rule
and Fourier image of~(\ref{norm})
\be
\label{CC}
\left\langle^{\phantom{`^{\phantom{`}}}}  C(X^i)(w)\partial_n
X^{i_1}(x_1)...\partial_n X^{i_l}(x_l)\right\rangle = \cr = {1\over
l!}{\partial^{~l} C\over \partial X^{j_1}...\partial
X^{j_l}}(X^i=0)\left\langle^{\phantom{`^{\phantom{`}}}} :X^{j_1}...X^{j_l}(w): \partial_n
X^{i_1}(x_{i_1})...\partial_n X^{i_l}(x_{i_l})\right\rangle = \cr = {1\over
l!}{\partial^{~l} C\over \partial X^{j_1}...\partial
X^{j_l}}(X^i=0) (-i \alpha' )^l {g^{i_1 j_1}...g^{i_l j_l}l!(w-\bar{w})^l\over
(x_1-w)(x_1-\bar{w})...(x_l-w)(x_l-\bar{w}) }\ee

Substituting last two formulae into (\ref{_action2}) and adding (\ref{cterm}) one obtains the
RR couplings to D-branes

\be
\label{_action3}
S=\sum_{m=0}^{\infty}\sum_{n=0}^{\infty}\sum_{r=0}^{\infty}
\int\limits d^{p+1} X {(i\pi \alpha')^{m+n+r} 2^r \over 32 (p'+1)!}
 \times \cr \times
Sp\left[\Gamma^{A_0}...\Gamma^{A^{\phantom{`}}_{p'}}\Gamma^{p+1}...\Gamma^9\Gamma^{11}
\Sigma^{\mu_1\nu_1}...\Sigma^{\mu_n\nu_n} \Sigma^{i_1 j
_1}...\Sigma^{i_m j_m} \Sigma^{\lambda_1 s_1}...\Sigma^{\lambda_r
s_r}\right] \times\cr\times {1\over n!m!r!} STr
\left\{C(X^\mu,X^i=|\lambda|\Phi^i)_{A_0...A_{p'}}F_{\mu_1\nu_1}(X^{\mu})...F_{\mu_n\nu_n}(X^{\mu})
\cdot  \right. \cr \left.   [\Phi_{i_1},\Phi_{j_1}](X^{\mu})...[\Phi_{i_m},\Phi_{j_m}](X^{\mu})
D_{\lambda_1}\Phi_{s_1}(X^{\mu})...D_{\lambda_r}\Phi_{s_r}(X^{\mu})^{\phantom{`^{\phantom{`}}}}  \right\} = \ee
\be
=\int d^{p+1} X^{\mu} STr
\left \{ \sum_{n=0}^{\infty} \sum_{m=0}^{[{p'+1 \over 2}]}\sum_{r=0}^{p'+1-2m}
{\lambda^m \over 2^m m!}
[\Phi_{i_1},\Phi_{j_1}]...[\Phi_{i_m},\Phi_{j_m}] \times \right. \cr \left.
\times
C(X^\mu,\Phi)_{j_1 i_1...j_m i_m k_1...k_r\sigma_{1}...\sigma_{s}} \times \right.
\cr \left. \times {\phantom{12}\lambda^{r^{\phantom{`^{\phantom{`^{\phantom{`}}}}}}} \over  r!} D_{\lambda_1}\Phi^{k_1}...D_{\lambda_r}\Phi^{k_r}  {\lambda^n \over 2^n n!}
F_{\mu_1\nu_1}...F_{\mu_n\nu_n} \right\}\times \cr
\times
\delta_{2n,p+2m-p'}\epsilon^{\lambda_1...\lambda_r\sigma_{1}...\sigma_{s}\mu_1\nu_1...\mu_n\nu_n},~~\cr ~s=p'+1-r-2m
\ee

Using the following notations
\be
\epsilon^{\mu_1...\mu_n}{\bf P}[C]_{\mu_1...\mu_n}=\epsilon^{\mu_1...\mu_n} \sum_{m=0}^{m=n}
\left({\lambda \over 2}\right)^n C_n^m D_{\mu_1}\Phi^{i_1}...D_{\mu_{m}}\Phi^{i_m}C_{i_1...i_m\mu_{m+1}...\mu_{n}
}, ~~m=n-k
\ee
and
\be
e^{I_{\Phi}I_{\Phi}}C=C_{A_0...A_{p'}}+\Phi^i\Phi^j C_{jiA_0...A_{p'-2}}
+{1\over 2}\Phi^i\Phi^j \Phi^{i'}\Phi^{j'} C_{ji{j'}{i'}A_0...A_{p'-4}}+...\ee
the final expression can be
represented as:
\be
S=\int  STr ~\left(
{\bf P} \left[^{\phantom{`^{\phantom{`}}}}e^{i \lambda  I_{\Phi} I_{\Phi}}C\left(X^\mu,X^i=|\lambda|\Phi^{i^{\phantom{`}}}\right)^{\phantom{`^{\phantom{`}}}}\right]\wedge e^{\lambda F^{\phantom{`^{\phantom{`}}}}}\right)_{top} \ee

\section{Conclusions}

\ \ \ In this paper we obtain low-energy effective action which describes the
interaction between massless Ramond-Ramond and gauge/matter fields by
calculating tree-level string scuttering amplitudes.
Many  results were already derived in this way.
Among them open-string low-energy action (so-called Born-Infeld) and it's
nonabelian generalizations \cite{Ts1,Ts2,Ts3,Pestun}.
This approach was already used in oder to check the results, derived in
other ways (see introduction).

Unfortunately  this approach usually is not useful   since the calculations
in all orders are very complicated.

Moreover after calculations it is necessary to carefully extract all   singularities
 (poles in external momentum) and this leads to additional difficulties. In the
general case this extractions could not be done before the explicit result
for string amplitudes is obtained.

Our goal was to find  the procedure, which extracts the poles before the explicit
result is obtained. We do not know how to prove our proposal  carefully and
leave it as assumption, but considering only the Wick pairings between vertexes,
 which never merge leads to correct result.

This assumption allow us to avoid tedious calculations, which ultimately difficult
to carry out in all orders. Really these calculations were successfully carried out    only in
leading orders \cite{Garo} for RR couplings.

It will be very interesting to prove this assumption and also generalize it to the case of arbitrary
fields. For instance the calculations with tachyon  could not be done using
the assumption in the described form because the off-shell amplitudes  (and therefor $SL_2$
non-invariant expressions) are relevant for this interactions \cite{RR1}.

We also derive the vertex operator for RR filed in the $(-{3\over 2},-{1 \over 2})$ picture.
It depends on RR fields themselves  rather than their fields strength and this simplify the derivation of low-energy action.
Note that this vertex is not BRST closed but it still leads to the right result inside the
correlators.

Thus we demonstrate that Myers-Chern-Simons action is really the consequence
of the superstring theory in the low-energy limit.

\section{Acknowledgments}

\ \ \ \ ~ Author  would like to thank    E.Akhmedov,
 A.Gerasimov and  V.Ch.Zhukovsky for initiating  this work,  useful
advises and numerous discussions.
 I am also grateful to A.Gorsky and V.Dolotin for reading the manuscript and useful remarks.

This work was partly supported  by the RFBR grant 01-02-17682a, RFBR grant
for Supporting of Young Scientists 00-02-06026, INTAS grant 00-334  and by the Russian
President's  grant 00-15-99296.

\section{Appendix A}
 In this appendix we are going to calculate the following integral

\be \label{I}
I=\int_{-\infty}^{\infty}dx_1...\int_{-\infty}^{\infty}dx_k
{(w-\bar{w})^k\over (x_1-w)(x_1-\bar{w})...(x_k-w)(x_k-\bar{w})}
\ee
 The most convenient way to do this is to make analytical
change of variables

\be x\rightarrow z={x-i \over
x+i},~~~~~x=i{1+z\over 1-z}
\ee
This is the transformation of the string world-sheet
from upper-half plane to the unit
disk. In fact, if $x\in R$ then $|z|=1$ and therefore $z=e^{it}$.
The constraint $+\infty>x_1>~...~>x_k>-\infty$ transforms to
$2\pi>t_1>...>t_k>0$. It is straightforward to check that

\be
dx={2idz\over(1-z)^2},~~~~~~~(x_1-x_2)=
{2i(z_1-z_2)\over(1-z_1)(1-z_2)} \ee so \be \label{II}
I=\oint_{|z_1|=1}dz_1...\oint_{|z_k|=1}dz_k
{(z_w-z_{\bar{w}})^k\over
(z_1-z_w)(z_1-z_{\bar{w}})...(z_k-z_w)(z_k-z_{\bar{w}})},\cr~\forall
i=\overline{1,..,k}~~~|z_i|=1
\ee
Since $|z_w|<1$ and $|z_{\bar{w}}|>1$ it is easy to evaluate
(\ref{II}) with the help of Cauchy theorem:

\be
I=(2 \pi i)^k
\ee
We see that $I$ does not depend on $w$. This is the consiquence of the
 $SL_2$ invariance of (\ref{I}) under

\be x\rightarrow x'=
{ax+b\over
cx+d},~~~~~~~(x_1-x_2)\rightarrow({x'}_1-{x'}_2)={(x_1-x_2)\over
(c x_1+d)(cx_2+d)}
\ee
Note that  $SL_2$ transforms the points
$+\infty$ and $-\infty$  into the same point and (\ref{I}) is
explicitly invariant under linear change of variables $x
\rightarrow x'=x+const$. The latter transformation belongs to $SL_2$ as well. All
these considerations allow us to put $w$ to any complex number in the upper half
plane without change  of the result.

\end{document}